\documentclass[aip,pra,twocolumn,showpacs]{revtex4}
\usepackage{graphicx,color}
\usepackage{amsmath}
\usepackage{longtable}
\begin{document}

\title{Accurate momentum transfer cross section for the attractive Yukawa potential}

\author{S. A. Khrapak\footnote{Also at Joint Institute for High Temperatures RAS, 125412 Moscow, Russia; Electronic mail: Sergey.Khrapak@dlr.de} }
\date{\today}
\affiliation{ Forschungsgruppe Komplexe Plasmen, Deutsches Zentrum f\"{u}r Luft- und Raumfahrt,
Oberpfaffenhofen, Germany}

\begin{abstract}
Accurate expression for the momentum transfer cross section for the attractive Yukawa potential is proposed.
This simple analytic expression agrees with the numerical results better than to within $\pm 2\%$  in the regime relevant for ion-particle collisions in complex (dusty) plasmas.
\end{abstract}

\pacs{52.20.Hv, 52.27.Lw}
\maketitle

The problem of classical scattering in the Yukawa (screened Coulomb) potential has been extensively investigated, mostly in the context of transport properties of ionized gases~\cite{Liboff,Mason,Lane}. More recently this topic has experienced renewed interest in view of its relevance to collisions between plasma and massive charged particles and, in particular, to the ion drag force in complex (dusty) plasmas~\cite{Northrop,Kilgore,id1,id2}.

The attractive Yukawa potential can be written in the form
\begin{equation}
U(r)=-\varepsilon(\lambda/r)\exp(-r/\lambda),
\end{equation}
where $\varepsilon$ and $\lambda$ are the energy and (screening) length scales. In the following $\lambda$ is used as the unit of length. Scattering in binary collisions can be conveniently characterized by the dimensionless scattering parameter $\beta=\varepsilon/mv^2$, which measures the strength of the interaction potential compared to the kinetic energy of colliding particles (here $m$ and $v$ are the reduced mass and relative velocity, respectively). It can be easily shown that the functional dependence of the scattering angle on the impact parameter $\chi(\rho)$ is governed by the single parameter $\beta$ and, therefore, the transport cross sections are functions of the scattering parameter $\beta$ only (for point-like particles).

Collisions in conventional electron-ion plasmas are normally characterized by quite small values of $\beta$. For the thermal velocity, the scattering parameter $\beta$ reduces to the plasma parameter (also referred to as the Debye nonideality parameter) $\beta=e^2/T\lambda=\Gamma_{\rm D}$~\cite{FIK}.
In the weakly coupled (ideal) plasma, where the binary collision approximation is meaningful, the plasma parameter is very small $\Gamma_{\rm D}\ll 1$. The scattering is mostly with small angles and the well known standard Coulomb scattering theory can be used to evaluate the transport cross sections.

In complex (dusty) plasmas the situation can be quite different due to high values of the particle charge. The typical values of the thermal scattering parameter for (attractive) ion-particle interactions ($\beta$ evaluated at an average kinetic energy of the ions) for micron-size particles in typical gas discharges are expected to lie in the range between $\simeq 1$ and $\simeq 30$\cite{MT}. Higher values, up to $\beta\simeq 70$, have been reported in experiments with big hollow microspheres of a diameter $\simeq 60$ $\mu$m~\cite{Nosenko, NosenkoCorr}. For sub-micron grains the values of $\beta$ below unity can also be realized. The purpose of this Brief Report is to propose a simple expression for the momentum transfer cross section in the attractive Yukawa potential, which is accurate to within $\pm 2\%$ in the wide range $0.1<\beta<100$, sufficient for the majority of practical situations in complex plasmas. This expression is superior to previous analytical approximations suggested in the literature, which demonstrate significant inaccuracies in the regime around $\beta\simeq 10$, where the momentum transfer cross section exhibits a pronounced non-monotonous behavior.

\begin{figure}
\includegraphics[width=8.5cm]{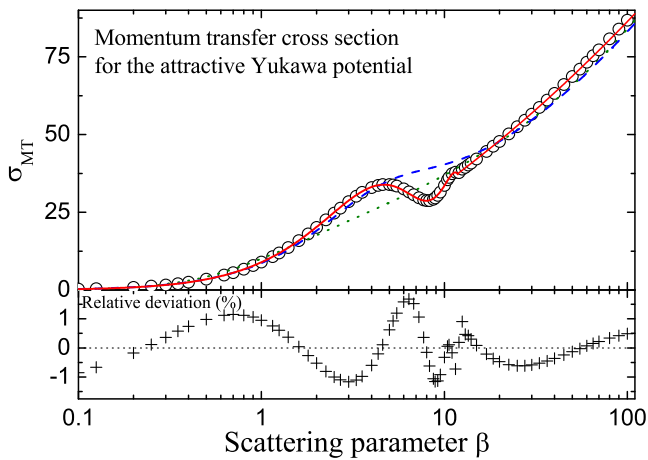}
\caption{(Color online) (a) Reduced momentum transfer cross section, $\sigma_{\rm MT}$, as functions of the scattering parameter $\beta$. Symbols correspond to the numerical results. The solid curve is the fit proposed in this paper. The dashed (dotted) curve is the fit from Ref.~\cite{Hutch} (Ref.~\cite{id3})  (b) Relative deviation of the present fit [Eq.~(\ref{Fit})] from the numerical data.
} \label{f1}
\end{figure}

The dependence of the scattering angle $\chi$ on the impact parameter $\rho$ and the resulting momentum transfer cross section
\begin{equation}
\sigma_{\rm MT}=2\pi\int_0^\infty [1-\cos\chi(\rho)]\rho d\rho
\label{crossection}
\end{equation}
have been evaluated numerically from  the conventional expressions~\cite{LL} in the very broad range of $\beta$ between $0.01$ and $10^6$, much broader than published previously in Refs.~\cite{id2,id3}. The results for the high-$\beta$ regime ($\beta\gg 1$) have been reported in Ref.~\cite{PRE2014}. Here the focus is on the intermediate transitional regime $0.1<\beta<100$.
The numerical results for the momentum transfer cross section pertaining to the range of $\beta$ under consideration are shown in Fig.~\ref{f1}(a) (here and throughout the paper, the cross section is given in units of $\lambda^2$). A remarkable property of the dependence $\sigma_{\rm MT}(\beta)$ is the non-monotonous behavior, with local maxima and minima, occurring in the region around $\beta\simeq 10$. The physics behind this behavior is the emergence of the barrier in the effective potential energy of colliding particles at $\beta\gtrsim \beta_{\rm cr}\simeq 13.2$~\cite{Kilgore,id2,FortovPR} (this can only occur for strongly {\it attractive} potentials). This barrier results in the divergence of the scattering angle at a certain ``transitional'' impact parameter $\rho_*$ for $\beta>\beta_{\rm cr}$, which plays a crucial role for the analysis of collisions and momentum transfer~\cite{id2,id3}. The non-monotonic behavior of $\sigma_{\rm MT}(\beta)$ is a consequence of the burification that the scattering angle $\chi(\rho)$ experiences on approaching $\beta_{\rm cr}$ (see for instance Fig. 16 from Ref.~\cite{FortovPR}).

Several expressions have been suggested in the literature to describe the dependence $\sigma_{\rm MT}(\beta)$ for the attractive Yukawa potential in the considered range of $\beta$. These are also plotted in Fig.~\ref{f1}. The dotted curve corresponds to Eq.~(9) from Ref.~\cite{id3}, the dashed line shows the fit by Eq.~(9) from Ref.~\cite{Hutch}. Both are reasonably accurate for small and large $\beta$, but can deviate from the numerical results by $\simeq 30\%$ in the intermediate regime $2\lesssim\beta\lesssim\beta_{\rm cr}$. Our main goal here is to propose an expression that can describe the non-monotonous behavior of $\sigma_{\rm MT}$ in the intermediate region more accurately. It would be useful to briefly remind first the main tendencies in the dependence $\sigma_{\rm MT}(\beta)$ in the low-$\beta$ and high-$\beta$ regimes.

In the low-$\beta$regime ($\beta\ll 1$) the ordinary Coulomb scattering theory is applicable. It assumes Coulomb potential and uses cutoff for impact parameters larger than the Debye length.
This results in the momentum transfer cross section $\sigma_{\rm C}=2\pi \beta^2\ln(1+1/\beta^2) \simeq 4\pi\beta^2\ln(1/\beta)$. This expression is correct only to the leading logarithmic term (the Coulomb logarithm). The additional to $\ln(1/\beta)$ terms of order ${\mathcal O}(1)$ can be obtained using more accurate consideration~\cite{Liboff,Kihara,Gould}.
A simple modification of the standard Coulomb scattering theory has been proposed in Ref.~\cite{id1} by re-defining the cutoff length. The modified cross section is
\begin{equation}\label{cs1}
\sigma_1(\beta)=4\pi\beta^2\ln(1+1/\beta).
\end{equation}
It reduces to the expression from the standard Coulomb scattering approach in the limit $\beta\ll 1$, but is applicable up to $\beta\sim {\mathcal O}(1)$.

In the opposite high-$\beta$  regime ($\beta>\beta_{\rm cr}$) the scattering angle exhibits the following properties~\cite{id2,id3,PRE2014}. For impact parameters below the transitional one $\rho_*$ scattering with large angles ($\pi<\chi< \infty$) occurs. The scattering angle grows monotonically until it diverges at $\rho=\rho_*$, orbiting trajectories are possible. For $\rho>\rho_*$ the scattering angle decreases rapidly due to exponential screening of the interaction potential. Interestingly, the dependence $\chi(\rho/\rho_*)$ exhibits quasi-universal behavior -- it is practically independent of $\beta$~\cite{id2,id3}. This implies that in this high-$\beta$ regime the cross section scales as $\sigma_{\rm MT}\propto \rho_*^2$. Taking into account that $\rho_*\propto \ln\beta$~\cite{id2,id3}, a simple two-term expression has been proposed recently~\cite{PRE2014}
\begin{equation}\label{cs2}
\sigma_2(\beta)=1.63\ln^2\beta+10.61\ln\beta.
\end{equation}
It agrees with the numerical results to within several percents in the range from $\beta_{\rm cr}$ to at least $10^6$.

Following the strategy of Ref.~\cite{Khrapak2014_2}, where accurate fits for the transport cross sections for the Lennard-Jones potential have been obtained, we use the scalings given by Equations (\ref{cs1}) and (\ref{cs2}) as the basis in the low-$\beta$ and high-$\beta$ regimes, respectively. These scalings are then multiplied by correction functions $f_1(\beta)$ and $f_2(\beta)$, to reach better agreement with the numerical results. The resulting expressions are then matched at some intermediate point around $\beta\simeq 10$. The low-$\beta$ correction function has the form
\begin{equation}\label{fit1}
f_1(\beta) =\sum_{i=0}^{4}c_i\beta^{i}.
\end{equation}
The appropriate coefficients are summarized in Table~\ref{Tab1}. In the high-$\beta$ regime it is sufficient to simply multiply $\sigma_2$ by a constant coefficient (close to unity), so that $f_2(\beta)\equiv 1.035$. The momentum transfer cross section is then
\begin{equation}\label{Fit}
\sigma_{\rm MT}(\beta) = \begin{cases} \sigma_1(\beta)f_1(\beta), & \beta< 11.871 \\ \sigma_2(\beta)f_2(\beta), & \beta>11.871 \end{cases}
\end{equation}
The two branches are matched at $\beta=11.871$, where $\sigma_{\rm MT}\simeq\sigma_1f_1\simeq \sigma_2f_2\simeq 37.50$.
The solid curve in Fig.~\ref{f1}(a) plots the fit by Eq.~(\ref{Fit}). The agreement with the numerical results is excellent, relative deviations are well within $\pm 2\%$, as documented in Fig.~\ref{f1} (b). Note that since the momentum transfer rate (needed for instance to calculate the ion drag force) results from the integration of this cross section with the velocity distribution function, it should be even more accurate than this.

\begin{table}[t!]
\caption{\label{Tab1} Fitting parameters entering equation (\ref{fit1}).}
\begin{ruledtabular}
\begin{tabular}{lllll}
$c_0$ &  $c_1$ & $c_2$ & $c_3$ & $c_4$   \\ \hline
$0.927956$  & $0.17968$  & $-0.09328$  & $0.01063$  & $-3.74479\times 10^{-4}$   \\
\end{tabular}
\end{ruledtabular}
\end{table}

It should be noted that we have considered the case of point-like particles interacting via the attractive Yukawa potential. In the context of ion-particle collisions in complex plasmas, the finite particle size can affect the magnitude of the momentum transfer cross section. It is relatively straightforward to take this effect into account as has been discussed in considerable detail previously~\cite{id2,MT,id3}. Therefore, we do not elaborate on this here.

To summarize, a simple analytical expression for the momentum transfer cross section for the attractive Yukawa potential is derived. This expression demonstrates much better accuracy, than those proposed previously. The accuracy is more than sufficient taking into account possible deviations from the Yukawa potential, the role of finite particle size, ion-neutral collisions and other effect which make real complex plasmas different from the ideal model considered in this work.

This work was partly supported by the Russian Foundation for Basic Research, Project No. 13-02-01099.

\end{document}